\documentstyle[prl,twocolumn,aps]{revtex}
\input{epsf}

\newcommand{\bq}{\begin{equation}}
\newcommand{\ee}{\end{equation}}
\newcommand{\fr}[2]{\frac{#1}{#2}}
\newcommand{\eps}{\varepsilon}

\begin{document}
\draft

\title{Localization in an imaginary vector potential}

\author{P.G.Silvestrov}

\address{Budker Institute of Nuclear Physics, 630090
Novosibirsk, Russia}

\maketitle

\begin{abstract}

Eigenfunctions of $1d$ disordered Hamiltonian with constant
imaginary vector potential are investigated. Even within the
domain of complex eigenvalues the wave functions are shown to be
strongly localized. However, this localization is of a very
unusual kind. The logarithm of the wave function at different
coordinates $x$ fluctuates strongly (just like the position of
Brownian particle fluctuates in time). After approaching its
maximal value the logarithm decreases like the square root of
the distance $\overline{( \ln|\psi_{max}/\psi| )^2} \sim
|x-x_0|$. The extension of the model to the quasi-$1d$ case is
also considered.

\end{abstract}


The non-Hermitean disordered Hamiltonians with imaginary vector
potential were introduced in Ref. \cite{Hatano} for the
description of vortex depinning from columnar defects by a
transverse magnetic field in superconductors (later the
analogous equations were used in the context of investigation of
some problems of biological growth \cite{Shnerb}). The features
of this unusual Hamiltonians have attracted the immediate and
wide interest
\cite{Efetov,Zee,Brouwer,Janik,Khorunzgenko,Brezin,Altland,Shnerb2}.
From the theoretical point of view the most exciting was the
prediction of the existence of a mobility edge in $1d$ in such
systems.  However, the main attention in the existing literature
was paid to the solution of eigenvalue problem in the imaginary
potential. The spectrum of this non-Hermitean Hamiltonian was
shown to consist of two segments of real eigenvalues and two
arcs of imaginary eigenvalues (see insert on the Fig. 1). The
domain of real energy was calculated in the initial paper
Ref.~\cite{Hatano}. The analytic formula relating the real and
imaginary part of complex eigenvalues for $1d$ model with weak
disorder was found in Ref.~\cite{Brouwer} (some other analytic
approaches and approximations were considered in Refs.
\cite{Zee,Janik,Khorunzgenko,Brezin}). While all eigenstates
with real energy turn out to be naturally localized, the complex
eigenvalues are usually associated with the extended states.
Nevertheless, in this note we show that even after the
transition to complex spectrum the wave functions remain
sufficiently localized in $1d$ imaginary vector potential. This
localization, however, is of a very unusual kind. The logarithm
of the wave function changes randomly with the increase of $x$.
The maximum of the wave function corresponds to the maximal
positive deviation of this random path from the initial
position. After approaching maximum the logarithm of the wave
function decreases also randomly being $\overline{(
\ln|\psi_{max}| - \ln|\psi(x)| )^2}
\sim |x-x_{max}|$. We will call the localization of the
eigenvectors of this type the Stochastic Localization.

The Schr\"{o}dinger equation for a particle in one
channel disordered ring with constant imaginary vector potential
has the form
\bq\label{Shred}
H\psi = -\left( \fr{\partial}{\partial x} -h\right)^2 \psi +
V(x) \psi = 
E\psi \ \ .
\ee
The boundary conditions are $\psi(x+L)=\psi(x)$ (the closed
ring). For wire with open ends the vector potential may be
washed out by the gauge transformation $\psi(x) \rightarrow
\exp(hx)\psi(x)$. Our results do not depend on the concrete
choice of the disordered potential $V(x)$ (it may be white noise
or something else). Also for analytic estimates we consider the
case of weak disorder. This means that both the wave length and
$h^{-1}$ are small compared to the mean free path.

We will consider only the states with large complex energy.
Eigenvectors with small real energy may be obtained by gauge
transformation from the usual localized states with $h=0$. The
most clear difference of the Hamiltonian Eq. (\ref{Shred}) from
the usual $1d$ disordered case is that at $h$ large enough the
left going and right going waves are no more divergent. The
corresponding energies for the momentum $\pm k$ are $E_{\pm}=k^2
-h^2 \pm 2ikh$. As a result one may look for the solution of 
Eq. (\ref{Shred}) of the form (analog of
the WKB method in Hermitean Quantum Mechanics \cite{ref})
\bq\label{exp}
\psi =\exp(i\sigma) \ \ .
\ee
The small corrections due to back
scattering may be estimated by the method which we use below
for the investigation of multichannel case.
The derivative expansion for $\sigma$  now gives
\begin{eqnarray}\label{WCB}
\sigma=\sigma_0 +\sigma_1 + ... \  &,& \\
\sigma_0= \int_0^x (\sqrt{E-V} -ih)dx' \  &;&
 \ \sigma_1 =\fr{i}{4} \ln(E-V) \ ... \ \ . \nonumber
\end{eqnarray}
The weak disorder may mix only the plain-wave states with
momentum $k'$ close to $k$. Therefore, only slow garmonics of
the potential $V_{kk'}$ came into play, which justifies the use
of WKB approximation \cite{WKB}. For our purposes it will be
enough to 
consider only the first term $\sigma_0$ of the expansion Eq.
(\ref{WCB}). The quantization condition now takes the form
$\sigma(L)-\sigma(0)=2\pi n$, which in the first order in $V$
leads to the energy $E=(k+ih)^2 +\langle V\rangle$ with $k=2\pi
n/L$ and $\langle V\rangle = L^{-1} \int_0^L Vdx$. The function
$\sigma(x)$ is almost real. However, due to a weak disorder it
acquires a small fluctuating imaginary part. Due to this
imaginary part the logarithm of the modulus of the wave function
(again in the leading order in $V$) behaves like
\bq\label{ln}
\ln |\psi(x)|^2 =\frac{h}{k^2+h^2} \int_0^x (V(x') -\langle
V\rangle) dx' \ \ .
\ee
Thus we see that our wave function may be large (exponentially!)
or small depending on the occasional sign of the integral. We
see also that the momentum $k$ emerges in the Eq. (\ref{ln})
only in the overall factor $(k^2+h^2)^{-1}$. As a result all
wave functions turn out to be localized at the same place. This
feature of localization in the strong imaginary potential is in
sharp contrast with what happens in the Hermitean $1d$
disordered Hamiltonians, where eigenfunctions with close energy
strongly repel each other in the coordinate space. In the case
of white noise disorder $\overline{V(x)V(x')}=D\delta(x-x')$ one
may find how fast the logarithm Eq. (\ref{ln}) changes from
point to point. Namely
\bq\label{ln2}
\overline{ (\ln |\psi(x)| - \ln |\psi(y)|)^2} =
\frac{h^2D|x-y|(L-|x-y|)}{4(k^2+h^2)^2 L} \ \ .
\ee
In particular this equation allows us to estimate, how the
wave function decreases after approaching the maximum. One may
introduce via the Eq. (\ref{ln2}) the typical size of the
wave packet (analog of the localization length)
\bq\label{ksi}
\xi_h \sim \fr{(k^2+h^2)^2}{h^2D} \ \ \ .
\ee
This localization length formally is of the same order of
magnitude as the usual Anderson localization length in $1d$
disordered wire $\xi_A \sim 1/D$. However, the nature of this
localization is completely different. In Anderson case all wave
functions decrease at large distances like $\exp (-x/\xi_A)$. In
our case the wave functions decrease like $\exp (-const
\sqrt{x/\xi_h})$ and even the constant in the argument of the
exponent fluctuates with $x$ and from sample to sample (no
thermodynamic limit).

Our results Eqs. (\ref{WCB}-\ref{ksi}) are valid only far from
the transition from real to complex spectrum in the limit of
weak disorder. In order to study the more general situation,
numerical simulations were performed. For numerical computations
it is natural to consider the tight binding variant of the model
(\ref{Shred})
\bq\label{tight}
-t\left( e^{-h} \psi_{n+1} + e^{h} \psi_{n-1}\right) + V_n
\psi_n= E\psi_n \ \ .
\ee
The approximate solution of this equation may be easily found
for the case $h\gg 1$ (compare also with \cite{Shnerb2})
\begin{eqnarray}\label{disk}
&& E_k\approx -t [ e^{h-i\phi} + e^{-h+i\phi}] \approx -t
e^{h-i\phi} \ ; \ \phi=2\pi \fr{k}{N} \\
&& \psi_n^{(k)} =\exp\left\{ in\phi +
e^{-h+i\phi}\sum_{m=0}^{n-1} (V_m -\langle V\rangle )
+O(e^{-2h}) \right\}  .\nonumber
\end{eqnarray}
Here integer $N$ and $k$ are the size of the closed chain and
the number of solution (momentum), $\langle V\rangle = N^{-1}
\sum_{n=0}^N V_n$. For concrete calculation we have used
$N=300$, $t=1$, $h=0.4$, and the random potential uniformly
distributed within the range $-1.5 <V_{n}< 1.5$. Thus we
consider the case of intermediate or even strong disorder. The
eigenvalues on the complex plane for one realization of disorder
are plotted on the insert on Fig.~1. The left segment of real
eigenvalues includes $29$ low energy states. The more
interesting for us are the squared wave functions
$|\psi^{(i)}_n|^2$ shown on the same figure (the wave functions
are normalized to unity in the usual sense $\sum |\psi_n|^2 =1$
and only the values $|\psi_n|^2>10^{-5}$ are shown). We have
shown the few typical states well below the phase transition
$i=9,10,11$ and the states around and above the transition
$i=27,30,36,50$. The solutions are numbered in accordance with
increase of $ReE$.  First, one sees that all the states below
and above the "mobility edge" are clearly localized. Second,
this localization has a very different form below and above the
transition. The consecutive real eigenfunctions $i=9,10,11$
practically do not overlap.  Alternatively, the states above the
transition (and even slightly below) turns out to be very
similar. This features of the wave functions are in explicit
accordance with what we expected from the analytic estimates for
weak disorder. On the Fig.~2 we have shown the
$\ln|\psi^{(i)}_n|^2$ for $i=4,50,150$.  As we have mentioned
before, up to exponentially small corrections the real wave
functions may be obtained from the solutions with $h=0$ via
$\psi_n(h)= e^{nh} \psi_n(h=0)$. In accordance with this
prediction we see on the Fig.~3 for $i=4$ that the logarithm of
the wave function decreases linearly from the maximum with
different slopes to the left and to the right.  Also as we
expected, the logarithm of the complex functions resembles to
large extent the Brownian path. Moreover, due to the Eq.
(\ref{disk}) at the middle of the zone ($\phi=\pi/4$) the
contribution of the leading order in $V$ changes only a phase of
the wave function but not an amplitude. The Eq.~(\ref{disk}) is
valid only for $h\gg 1$ and this peculiarity at the center of
the band should not necessary survive in our simulations.
However, as we see, the effect of localization is strongly
suppressed (but is not washed out completely!) for $i=150$.

In order to introduce the analog of orthonormal basis in
non-Hermitean case one has to consider two sets of eigenvectors:
$H\psi_R=E\psi_R$ and $\psi_L^{\ T} H=E\psi_L^{\ T}$. Up to now
in fact we have considered only the $\psi_R$. For our specific
choice of the Hamiltonian Eq.~(\ref{Shred}) evidently
$\psi_L(E,h) \equiv\psi_R(E^*,-h)$. Moreover, in the leading
approximation, with only $\sigma_0$ left in the Eq.~(\ref{exp}),
one simply has $\psi_L(x) \psi_R(x) = const$. Thus we see that
the growth (exponential!) of $\psi_R$ at the maximum is
compensated by the decrease of $\psi_L$. This compensation is
not an artifact of approximation Eqs.~(\ref{exp},\ref{WCB}). For
example, following the authors of the Ref.~\cite{Hatano}, one
may introduce the conserving current $J= \psi_L' \psi_R -\psi_L
\psi_R' +2h\psi_L \psi_R \ ; \ dJ/dx =0$. This means that the
strong effect of stochastic localization may be washed out if
the observables are bilinear in $\psi_R$ and $\psi_L$, but
should be seen if the observables depend on only one $\psi_R$ or
$\psi_L$.  Surprisingly one finds the examples of both kinds
within the most popular application of non-Hermitean Quantum
Mechanics.  Namely, the authors of Refs.~\cite{Hatano,Vinokur}
have used the Hamiltonian Eq.~(\ref{Shred}) for description of
Abrikosov vortex depinning by transverse magnetic field in the
hollow superconducting cylinder with random columnar defects.
The probability to find vortex at the point $x$ on the
transverse slice $\tau$ of the cylinder with length $L_{\tau}$
was shown to have the form $P(x,\tau) \equiv Z^{-1} \langle
\Psi^f| e^{-(L_{\tau}-\tau)H} |x\rangle \langle x|e^{-\tau
H}|\Psi^i\rangle$. For very large $L_{\tau}$ only the ground
state contribution survives in the evolution operator
$e^{-L_{\tau}H}$. In this case naturally \cite{Hatano,Tauber}
$P(x,\tau) \sim \psi_R (x)\psi_L(x)$ for $\tau,L_{\tau}-\tau
\sim L_{\tau}$. We see that the effect of localization above
depinning transition, which we consider in this paper, could
hardly be seen inside the cylinder \cite{thanks}. However, at
the top of cylinder $P(x,L_{\tau}) \sim \psi_R (x)$ and at the
bottom $P(x,0) \sim \psi_L(x)$ the "stochastic localization",
which we have introduced, turns out to be of $100\%$ importance.
We leave the more detailed discussion of the localization above
depinning in case of many vortices for a separate publication
\cite{Silv}.

Up to now we have studied the effect of imaginary vector
potential only in the $1d$ systems. Consider now the closed
strip of the width $l$ and length $L$ with random weak disorder
$\overline{V(x,y)V(x',y')} = D_2\delta(x-x')\delta(y-y')$ and
imaginary potential along the ring (the 
Hamiltonians of the kind Eq.~(\ref{Shred}) in more than $1d$
have been also considered in Refs.
\cite{Hatano,Altland,Zee2,Hatano2}). The Hamiltonian for the 
strip differs from that of the Eq.~(\ref{Shred}) by the only
additional term $-\partial^2/\partial y^2$. It is convenient to
look for the solutions of this Schr\"{o}dinger equation of the
form
\bq\label{sol}
\psi(x,y) =\sum \psi_n(x)\sqrt{{2}/{l}}\sin(q_ny) \ \ ,
\ee
with $q_n=n\pi /l$. The analog of the Eq.~(\ref{Shred}) now reads
\bq\label{Shredy}
-\left( {\partial}/{\partial x} -h\right)^2 \psi_n + \sum_m
V_{nm} \psi_m = 
\eps_n\psi_n \ \ ,
\ee
where the longitudinal energy $\eps_n=E-q_n^2$. In the case of
white noise disorder $\overline{V_{nn}(x)V_{nn}(x')} =
(3D_2/2l)\delta(x-x')$ and for $m$ and $n$ different
$\overline{V_{nm}(x)V_{mn}(x')} = (D_2/l)\delta(x-x')$. At least
for small $l$ (or large $\eps_n-\eps_m$) and for disorder weak
enough one may consider the hopping $V_{nm}$ in Eq.
(\ref{Shredy}) as a perturbation. The zeroth approximation for
the wave function in this simple case is again
$\psi_n^{(0)}=\exp \{\int^x (\sqrt{\eps_n -V_{nn}}-ih)dx'\}$
with some given $n$ and $E=\eps_n +q_n^2 = (k+ih)^2 +q_n^2$.
Physically interesting, however, is the range of validity of
this single chanel approximation. The corrections to
$\psi_n^{(0)}$ are described by the equation
\bq\label{Shred2}
[-\left( {\partial}/{\partial x} -h\right)^2 - \eps_m]
\psi_m^{(1)} = -V_{mn} \psi_n^{(0)} \ \ .
\ee
The solution of this equation may be found by the Green's
function method. Let for definiteness $\eps_m-\eps_n >0$ (note
that $Im \eps_m\equiv Im\eps_n$). In this case the Green's
function for the Eq. (\ref{Shred2}) $G(x-y)$ equals zero at
$x>y$ and has a cusp at $x=y$. The solution reads
\bq\label{sol1}
\psi_m^{(1)}(x)= \int_x^{\infty} \fr{e^{\lambda_1(x-y)}-
e^{\lambda_2(x-y)}}{\lambda_2-\lambda_1} V_{mn}(y)
\psi_n^{(0)}(y) dy \ ,
\ee
where $\lambda_1\approx 2h-ik$ and $\lambda_2\approx ik
+(\eps_m-\eps_n)/(h-ik)$ ($k$ is related with $E$ as we have
shown above and we suppose that $|\eps_m-\eps_n|\ll \eps_n$).
For $|n-m|\ll n$ one has $Re \lambda_2 \ll k,h$. However, if
$(k^2+h^2) |\eps_m-\eps_n|\gg hD_2/l$ (compare with Eq.
(\ref{ksi})) the amplitude of the exponent $|\exp(\lambda_2 x)|$
still vary with $x$ much faster than $|\psi_n^{(0)}(x)|$. In
this case the averaged value 
\bq\label{aver}
\overline{|\psi_m^{(1)}|^2} = \fr{D_2}{32\pi hq} \fr{1}{n-m}
|\psi_n^{(0)}(x)|^2 \ \ .
\ee
The analogous formula for $\eps_m-\eps_n <0$ is obtained after
the replacement $n-m \rightarrow |n-m|$.
In the similar way one may find the correction to the direct
potential $V_{nn}$ induced by the hopping $V_{nm}$ in the second
order of perturbation theory. This correction in its turn
renormalizes the speed of stochastic growth of the amplitude of
wave function Eq. (\ref{ln2}) and the localization length  Eq.
(\ref{ksi}). The generalization of the Eq. (\ref{ln2}) for the
strip gives
\begin{eqnarray}\label{gen}
&& \overline{(\ln|\psi_n^{(0)}(x)/\psi_n^{(0)}(y)|)^2}= \\
&& \ \ \ \ \fr{h^2|x-y|D_2}{4(k^2+h^2)l} \left\{ \fr{3}{2}  
+\fr{D_2}{32\pi
hq} \sum_{m\ne n} \fr{1}{|n-m|} +...\right\} . \nonumber 
\end{eqnarray}
Here $|x-y|\ll L$. The sum over $m$ in the r.h.s. is
logarithmically divergent. The summation should be cut at
$|n-m|\sim kl$ (we suppose that $k\sim q\sim h$). Thus we see
that at small $l$ the growth of the wave function simply
reproduces that of the single chanel case Eq.~(\ref{ln2}) with
the effective localization length proportional to the
width of the strip $\xi_h \sim l$. Only at $D_2 \ln(kl)/hq \sim
1$ the features of wave functions in a strip became sufficiently
different from those in $1$-chanel wire. One may interpret this
results as an indication of the existence of exponentially large
localization length on $2d$ plane with imaginary vector
potential. 

One may easily repeat the calculations
Eqs.~(\ref{sol}-\ref{gen}) also in the case of thick $3d$ wire.
The result of such generalization will be the formula analogous
to the Eq.~(\ref{gen}) and $\xi_h \sim l^2/D_3$ ($l$ is the
typical transverse size of the wire). Also like in $2d$ case the
new physics starts only at exponentially thick wire $D_3
\ln(kl)/q \sim 1$. However, the concrete behaviour of the
perturbed wave function may be completely different for $2d$ and
$3d$. In a strip due to a complex energy $E=(k_{\|}+ih)^2
+q_{\bot}^2$ there is no degeneracy of different transverse
channels. Thus the only new phenomena we may expect with
increase of the width of the strip is localization of wave
functions in transverse direction due to the mixing of channels
with very close $q_{\bot}$. In $3d$ case the analog of the
Eq.~(\ref{aver}) describes the diffusive mixing of many channels
with close $(\vec{q}_{\bot})^{2}$ but completely different
$\vec{q}_{\bot}$.

In summary, we have shown in this paper that the transition from
real to complex spectrum in $1d$ disordered systems with
imaginary vector potential is not a delocalization transition.
However, the nature of localization below and above the
transition is completely different. The most interesting is the
localization at complex energies, there the wave function
decreases (increases) like the exponent of $\sqrt{x/\xi_h}$ and
this exponent even has no a well defined thermodynamic limit. As
for physical applications, the localization at complex energies
becomes invisible for observable values bilinear in left- and
right- eigenvectors of non-Hermitean Hamiltonian, but is of
100\% importance for observables depending on only one of them
$\psi_L$ or $\psi_R$.  For example, the effect which we consider
will lead to the strong modification of the distribution of flux
lines at the ends of superconducting cylindrical shell with
random columnar defects.  For the $2d(3d)$ generalizations of
the model the same effect takes place with the effective
localization length $\xi_h$ proportional to the transverse
size(area) at least until the system became exponentially thick.

Author is thankful to L.~F.~Khailo for help in computations. The
work was supported by RFBR, grant 98-02-17905.


\vspace{-1cm}
\begin{figure}[t]
\epsfxsize=9cm
\epsffile{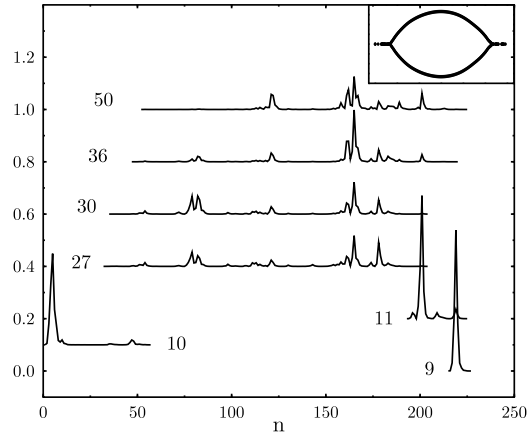}
\vglue 0.2cm
\medskip
\caption{The
$|\psi^{(i)}_n|^2$ for the model Eq.~(7) for 
states with numbers $i=9,10,11,27,30,36,50$. We use $N=300$,
$t=1$, $h=0.4$, and $-1.5 <V_{n}< 1.5$. The first
vector with complex energy is $i=30$. All states are clearly
localized. However, the states far below the "mobility edge"
differ strongly from each other, while the complex states look
quite similar. Insert shows the eigenvalues on the complex plane.
}
\end{figure}

\vspace{-1cm}
\begin{figure}[t]
\epsfxsize=9cm
\epsffile{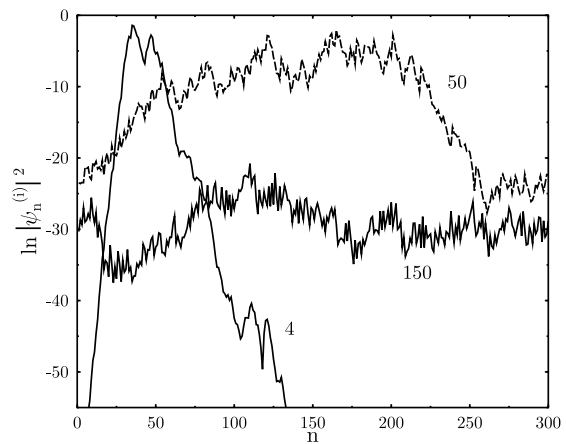}
\vglue 0.2cm
\medskip
\caption{The $\ln|\psi^{(i)}_n|^2$ for $i=4$
("localized") and $i=50,150$ ("stochastic") states. The data
for $i=150$ was multiplied by the factor $5$. The Brownian
curves for complex states and different left and right slope for
$i=4$ are clearly seen.
}
\end{figure}


\begin{thebibliography}{99}
\bibitem{ref} The WKB approximation for investigation the
spectrum of non-Hermitean Hamiltonians of the kind (\ref{Shred})
was first used in Refs. \cite{Brezin,Zee2}.
\end{thebibliography}
\end{document}